\documentclass[pre,bibtex,email,twocolumn,showpacs,showkeys,preprintnumbers,amsmath,amssymb,nofootinbib]{revtex4-1}

\usepackage{amsmath}
\usepackage{latexsym}
\usepackage{appendix}

\def\[{\left\lbrack}
\def\]{\right\rbrack}

\def\({\left(}
\def\){\right)}

\newcommand{\be}{\begin{equation}}
\newcommand{\ee}{\end{equation}}
\newcommand{\ea}{\end{eqnarray}}
\newcommand{\ba}{\begin{eqnarray}}

\begin{document}

\title{Extending the D'Alembert Solution to Space-Time Modified
Riemann-Liouville Fractional Wave Equations}

\author{Cresus F. L. Godinho$^{a}$}
\email{crgodinho@ufrrj.br} 
\author{J. Weberszpil$^{b}$}
\email{josewebe@ufrrj.br}
\author{J. A.
Helay\"{e}l-Neto$^{c}$}
\email{helayel@cbpf.br}
\affiliation{$^a$Grupo de F\' isica Te\'orica, 
Departamento de F\'{\i}sica, Universidade Federal Rural do Rio de Janeiro,\\
BR 465-07, 23890-971, Serop\'edica, RJ, Brazil\\
$^b$Universidade Federal Rural do Rio de Janeiro, UFRRJ-IM/DTL\\
 Av. Governador Roberto Silveira s/n- Nova Igua\c{c}\'u, Rio de Janeiro,
Brasil
\\
$^c$Centro Brasileiro de Pesquisas F\' isicas (CBPF),
Rua Dr. Xavier Sigaud 150, Urca, \\
22290-180, Rio de Janeiro, Brazil \\\\
\today\\}


\begin{abstract}
\noindent In the realm of complexity, it is argued that adequate modeling of
TeV-physics demands an approach based on fractal operators and fractional
calculus (FC). Non-local theories and memory effects are connected
to complexity and the FC. The non-differentiable nature of the microscopic
dynamics may be connected with time scales. Based on the Modified
Riemann-Liouville definition of fractional derivatives, we have worked
out explicit solutions to a fractional wave equation with suitable
initial conditions to carefully understand the time evolution of classical
fields with a fractional dynamics. First, by considering space-time
partial fractional derivatives of the same order in time and space,
a generalized fractional D'Alembertian is introduced and by means
of a transformation of variables to light-cone coordinates, an explicit
analytical solution is obtained. To address the situation of different
orders in the time and space derivatives, we adopt different approaches,
as it will become clear throughout this paper. Aspects connected to
Lorentz symmetry are analyzed in both approaches.
\end{abstract}

\maketitle

\pagestyle{myheadings}
\markright{\it Extending the D'Alembert Solution to Space-Time Modified
Riemann-Liouville Fractional Wave Equations}

\newpage

\section{Introduction}
During the last decade the interest of physicists in non-local field
theories has been steadily increasing. The main reason for this development
is the expectation, that the use of these field theories will lead
to a much more ele\-gant\- and effective way of treating problems in particle
and high-energy physics, as it is possible up to now with local field
theories. A particular subgroup of non-local field theories is described
with operators of fractional nature and is specified within the framework
of fractional calculus. Fractional calculus provides us with a set
of  axioms\- and methods to extend the concept of a derivative operator
from integer order n to arbitrary order $\alpha$, where $\alpha$
is a real value. Problems involving fractional integrodifferential
is an attractive framework that in recent years awakened the interest
of some researchers in the study on the fractional dynamics in many
fields of physics such as in anomalous diffusion\cite{Klafter-Metzler},
mechanics, engineering and other areas \cite{Jun-Sheng Duan,Goldfain1-Frac Dyn Standard Model}
and to deal with complex systems.

We can describe a complex system, as an \textquoteleft{}open\textquoteright{}
system involving \textquoteleft{}nonlinear interactions\textquoteright{}
among its subunits which can exhibit, under certain conditions, a
marked degree of coherent or ordered behavior extending well beyond
the scale or range of the individual subunits. Usually it consists
of a large number of simple members, elements or agents, which interact
with one another and with the environment, and which have the potential
to generate qualitatively "new" collective behavior, the manifestations
of this behavior being the spontaneous creation of new spatial, temporal,
or functional structures. Within the realm of complexity, new questions
in fundamental physics have been raised, which cannot be formulated
adequately using traditional methods. Consequently a new research
area has emerged, which allows for new insights and intriguing results
using new methods and approaches.

The interest in fractional wave equations aroused in 2000,
with a publication by Raspini \cite{Raspini 2000}. He demonstrated
that a 3-fold factorization of the Klein-Gordon equation leads to a 
fractional Dirac equation which contains fractional derivative operators
of order $\alpha\,=\,2/3$ and, furthermore, the resulting $\gamma$
- matrices obey an extended Clifford algebra \cite{Herrman1}, 
let us quote that, in the same year, the integer case was also 
studied in the work of Ref. \cite{Misha}.

It is well known that most of physical systems might be described
by Lagrangian functions and that frequently their dynamical variables
are on first order derivatives. This occurs probably because classical
and quantum theories with superior order derivatives do not present
a lower limit to the energy. Besides, the quantum field theories containing
superior derivatives of fields usually present states with negative
norm or ghost states and consequently a costly price is payed with
the lost of unitarity violation \cite{Fa}.

The quantum dynamics of a system can be described by operators acting
on a state vector, this is a good reason to study operators algebras
and its unfolds. The language of operators is very useful and important
to the physicists, we know today that the quantum analog of derivatives
like, for instance, $\partial/\partial t$ and $\partial/\partial x$
are operators. We still know that derivatives with respect to $q_{k}$
and $p_{k}$ can be represented by equations like Poisson brackets
and its quantum analog are commutators involving operators. The modern
theory of pseudo-differential operators took its shape in the sixties,
but we can consider the thirties, because the quantization problem,
preliminarily solved by Weyl, and since the 1980's this tool has also
yielded many significant results in nonlinear partial differential equations
PDE.

Recently, Guy Jumarie \cite{Jumarie1} proposed a simple alternative
definition to the Riemann\textendash{}Liouville derivative. His modified
Riemann\textendash{}Liouville derivative has the advantages of both
the standard Riemann\textendash{}Liouville and Caputo fractional derivatives:
it is defined for arbitrary continuous (non differentiable) functions
and the fractional derivative of a constant is equal to zero. Jumarie's
calculus seems to give a mathematical framework for dealing with dynamical
systems defined in coarse-grained spaces and with coarse-grained time
and, to this end, to use the fact that fractional calculus appears
to be intimately related to fractal and self-similar functions.

We would like to stress that the choice of \smallskip Jumarie's approach, 
besides the points already mentioned, is justified by the fact that chain 
and \-Leibinitz rules acquires a simpler form, which helps a great deal if changes of 
coordinates are performed. Moreover, causality seems to be more easily obeyed in a 
field-theoretical construction if we adopt Jumarie's approach.

As pointed out by Jumarie, non-differentiability and randomness are
mutually related in their nature, in such a way that studies in fractals
on the one hand and fractional Brownian motion on the other hand are
often parallel in the same paper. A function which is continuous everywhere
but is nowhere differentiable necessarily exhibits random-like or
pseudo-random-features, in the sense that various samplings of this
functions on the same given interval will be different. This may explain
the huge amount of literature which extends the theory of stochastic
differential equation to stochastic dynamics driven by fractional
Brownian motion.

The most natural and direct way to question the classical framework
of physics is to remark that in the space of our real world, the generic
point is not infinitely small (or thin) but rather has a thickness.
A coarse-grained space is a space in which the generic point is not
infinitely thin, but rather has a thickness; and here this feature
is modeled as a space in which the generic increment is not dx, but
rather $(dx)^{\alpha}$and likewise for coarse grained with respect
to the time variable t.

It is noteworthy, at this stage, to highlight the interesting work by 
Nottale \cite{Nottale} , where the notion of fractal space-time is first introduced.

In this work we claim that the use of an approach based on a sequential
form of Jumarie's Modified Riemann-Liouvile \cite{Jumarie1} is adequate
to describe the dynamics associated with fields theory and particles
physics in the space of non-differentiable solution functions or in
the coarse-grained space-time. Some possible realizations of fractional
wave equations are given and the proposed solutions are analyzed.
Based on this approach, we have worked out explicit solutions to a
fractional wave equation with suitable initial conditions to carefully
understand the time evolution of classical fields with a fractional
dynamics. This has been pursued in (1+1) dimensions where the adoption
of the light-cone coordinates allow a very suitable factorization
of the solution in terms of left-and-right-movers. First, by considering
space-time partial fractional derivatives of the same order in time
and space, a generalized fractional D'alembertian is introduced and
by means of a transformation of variables to light-cone coordinates,
an explicit analytical solution is obtained. Next, we address to the
problem of different orders for time and space derivatives. In this
situation, two different approaches have been adopted: one of them
takes into account a non-differentiable space of solutions, whereas
the other one considers a coarse-grained space-time as non-differentiable.
For the former, there emerges an indicative of chiral symmetry violation.
The latter points to a solution that depends on both the space and
time orders of the derivatives. Aspects of Lorentz transform and invariance
conditions are also analyzed. It is important to note that we are
not assuming the validity of the semi-group property of the fractional
derivatives \cite{Raspini 2000} and are not working with generalized
functions in the distributions sense \cite{Gelfand} nor taking fractional
power of operator \cite{Giambiagi}. Also, the solution technique
here proposed does not make uses of Fourier transform and not necessarily
Laplace transform. The work is organized as follows: After this Introduction,
the Jumarie's modified fractional derivatives are briefly presented
in Section 2. In the Section 3, the fractional wave equation is presented
in the coarse-grained space-time and in non-differentiable function
space of solutions. In Section 4 Lorentz transform and invariance
conditions are analyzed. Following, in Section 5 we present an example.
Finally, in Section 6, we cast our Concluding Comments. Two Appendices
follow. The Appendices A and B treat respectively the aspects of calculation
for different exponents and details of Lorentz transforms.\vfill

\section{Jumarie's fractional calculus and Modified Riemann-Liouville fractional
derivative}

The well-tested definitions for fractional derivatives, so called
Riemann-Liouville and Caputo have been frequently used for several
applications in scientific periodic journals. In spite of its usefulness
they have some dangerous pitfalls. For this reason, recently it was
proposed another definition for fractional derivative \cite{Jumarie1},
so called Modified Riemann-Liouville (MRL) fractional derivative and
its basic definition is

\begin{eqnarray}
D^{\alpha}f(x) & = & {\displaystyle \lim_{x\rightarrow0}\, h^{-\alpha}\sum_{k=0}^{\infty} \,(-1)^{k}\,{\alpha \choose k} f\left(x+(\alpha\,-\, k)h\right)}=\nonumber \\
 & = & \frac{{1}}{\Gamma(1-\alpha)}\frac{d}{{dx}}\int_{0}^{x}(x\,-\, t)^{-\alpha}\left(f(t)-f(0)\right)dt; \nonumber \\
0 & <& \alpha<1.\label{eq:1}
\end{eqnarray}

Some advantages can be cited, first of all, using the MRL definition
we found that derivative of constant is zero, and second, we can use
it so much for differentiable as non differentiable functions. They
are cast as follows:
\paragraph*{(i) Simple rules:}

\begin{eqnarray}
D^{\alpha}K &=&0, \nonumber \\ 
Dx^{\gamma}&=&\frac{\Gamma(\gamma+1)}{\Gamma(\gamma+1-\alpha)}x^{\gamma-\alpha},\;\gamma>0, \nonumber \\
(u(x)v(x))^{(\alpha)} &=&u^{(\alpha)}(x)v(x)+u(x)v^{(\alpha)}(x).
\end{eqnarray}

\paragraph*{(ii) Simple Chain Rules:}

\begin{equation}
\frac{d^{\alpha}}{dx^{\alpha}}f[u(x)]=\frac{d^{\alpha}f}{du^{\alpha}}\,\left(\frac{du}{dx}\right)^{\alpha}.\label{eq:Chainrule nondif func-1}
\end{equation}
For non differentiable functions

\begin{equation}
\frac{d^{\alpha}}{dx^{\alpha}}f[u(x)]=\frac{df}{du}\,\frac{d^{\alpha}u}{dx^{\alpha}}.\label{eq:chain rule space-time coarse-1}
\end{equation}
for coarse-grained space-time.

For more details, the readers can follow the Ref. \cite{Jumarie Book} which contains all the basic for the formulation 
of a fractional differential geometry in coarse-grained space, and refers to an extensive use of coarse-grained phenomenon.

Now that we have set up these fundamental expressions, we are ready
to carry out the calculations of main interest: the solutions to our
fractional wave equations

\section{Fractional wave equation in the MRL sense}

Since the semi group properties for fractional derivatives in general
does not hold, we used the Miller-Ross sequential derivative \cite{Miller-Ross}
in the MRL sense. Incidentally, the Miller-Ross sequential derivative is a systematic 
procedure that carries out a fractional higher-order derivative while avoiding the 
recursive application of many single derivatives taken after each other.
Moreover, we took the option to carry out the sequence of derivatives in the 
cascade form, in MRL sense, as done in the work of Ref. \cite{Jumarie2}.

\subsection{Coarse-grained space-time}

Let us Suppose a class of transformations that maps from the variables
x',t' to $u=\gamma x'^{\alpha}+\lambda\gamma t'^{\beta}$to illustrate
our prescription to apply the chain rule. Notice that the dimension
of $\lambda-parameter$ is such that $\lambda t'^{\beta}$matches
its dimension with $x'^{\alpha}$.

Performing successive derivatives and using the chain rule \eqref{eq:chain rule space-time coarse-1},
we can write

\begin{equation}
\frac{\partial^{\alpha}\phi}{\partial x'^{\alpha}}=\frac{d\phi}{du}\frac{\partial^{\alpha}u}{\partial x'^{\alpha}}=\frac{d\phi}{du}\gamma\Gamma(\alpha+1),
\end{equation}

\begin{equation}
\frac{\partial^{\alpha}\partial^{\alpha}\phi}{\partial x'^{\alpha}\partial x'^{\alpha}}=\frac{d^{2}\phi}{du^{2}}\gamma^{2}\Gamma^{2}(\alpha+1),\label{eq:d2xalfa}
\end{equation}

\begin{equation}
\frac{\partial^{\beta}\phi}{\partial t'^{\beta}}=\frac{d\phi}{du}\frac{\partial^{\beta}u}{\partial t'^{\beta}}=\frac{d\phi}{du}\lambda\gamma\Gamma(\beta+1),
\end{equation}

\begin{equation}
\frac{\partial^{\beta}\partial^{\beta}\phi}{\partial t'^{\beta}\partial t'^{\beta}}=\frac{d^{2}\phi}{du^{2}}\lambda^{2}\gamma^{2}\Gamma^{2}(\beta+1).\label{eq:d2tbeta}
\end{equation}

Substituting Eq.\eqref{eq:d2xalfa} into the Eq. \eqref{eq:d2tbeta}
yields:

\begin{equation}
\frac{\partial^{\alpha}}{\partial x^{\alpha}}\frac{\partial^{\alpha}}{\partial x^{\alpha}}\phi\left(x,t\right)-\frac{\Gamma^{2}(\alpha+1)}{\lambda^{2}\Gamma^{2}(\beta+1)}\frac{\partial^{\beta}}{\partial t^{\beta}}\frac{\partial^{\beta}}{\partial t^{\beta}}\phi\left(x,t\right)=0.\label{eq:wave equation com labda e gamas}
\end{equation}

\subsection{Non-differentiable functions}

Supposing a class o transforms that maps from x',t' to $u=\gamma x'+v\gamma t'$
and using the chain rule \eqref{eq:Chainrule nondif func-1}, we can
write

\begin{equation}
\frac{\partial^{\alpha}\phi}{\partial x'^{\alpha}}=\frac{d\phi}{du}\frac{\partial^{\alpha}u}{\partial x'^{\alpha}}=\gamma\frac{d^{\alpha}\phi}{du^{\alpha}}\left(\frac{\partial u}{\partial x'}\right)^{\alpha}=\gamma\frac{d^{\alpha}\phi}{du^{\alpha}},
\end{equation}
and

\begin{equation}
\frac{\partial^{\alpha}\partial^{\alpha}\phi}{\partial x'^{\alpha}\partial x'^{\alpha}}=\gamma^{2}\frac{d^{\alpha}}{du^{\alpha}}\frac{d^{\alpha}\phi}{du^{\alpha}}.
\end{equation}

Similarly, for a fractional time derivative, we have

\begin{equation}
\frac{\partial^{\beta}\phi}{\partial t'^{\beta}}=\frac{d^{\beta}\phi}{du^{\beta}}\left(\frac{\partial u}{\partial t'}\right)^{\beta}=\gamma\frac{d^{\beta}\phi}{du^{\beta}}\left(\pm v\right)^{\beta},
\end{equation}
 and

\begin{equation}
\frac{\partial^{\beta}}{\partial t'^{\beta}}\frac{\partial^{\beta}\phi}{\partial t'^{\beta}}=\gamma^{2}\frac{d^{\beta}}{du^{\beta}}\frac{d^{\beta}\phi}{du^{\beta}}\left(\pm v\right)^{2\beta}.
\end{equation}

The connection can be done for $\alpha=\beta,$which leads to

\begin{equation}
\frac{\partial^{\alpha}}{\partial x^{\alpha}}\frac{\partial^{\alpha}}{\partial x^{\alpha}}\phi\left(x,t\right)-\frac{1}{v^{2\alpha}}\frac{\partial^{\alpha}}{\partial t^{\alpha}}\frac{\partial^{\alpha}}{\partial t^{\alpha}}\phi\left(x,t\right)=0.
\end{equation}

Now with $t$ and $x$ with the same power, $v$ has the dimension
of speed.

\subsubsection{Fractional D'alembertian of same space-time partial derivative order}

The D'alembertian can now be established as:

\begin{equation}
\frac{\partial^{\alpha}}{\partial x^{\alpha}}\frac{\partial^{\alpha}}{\partial x^{\alpha}}\phi\left(x,t\right)-\frac{1}{v^{2\alpha}}\frac{\partial^{\alpha}}{\partial t^{\alpha}}\frac{\partial^{\alpha}}{\partial t^{\alpha}}\phi\left(x,t\right)=0,\quad0<\alpha<1\label{eq:Frac Dal same space-time order}
\end{equation}

Proceeding now a variable change in a light cone, right and left movers,
respectively, we can write

\begin{equation}
\left\{ \begin{array}{cc}
\xi=x-vt\\
\eta=x+vt
\end{array}\right.,\label{eq:left-right mover}
\end{equation}

Assuming a non-differentiable space of solutions, the chain rule 
in the MRL sense\eqref{eq:Chainrule nondif func-1} and using Eqs.
\eqref{eq:Frac Dal same space-time order}, \eqref{eq:left-right mover},
after some algebraic manipulations, we obtain

\begin{equation}
{\displaystyle \frac{\partial^{\alpha}}{\partial\xi^{\alpha}}\frac{\partial^{\alpha}}{\partial\eta{}^{\alpha}}\phi\left(\xi,\eta\right)}=0.
\end{equation}

In connection with wave equations modified to more complex structures, 
we would like to call attention to the work of Ref. \cite{Nobre} where the authors 
propose non linear wave equations with modified functions.

\subsection*{\label{sub:Solution-into-the light cone}Solutions in the light-cone
coordinates:}

The form of the Eq. \eqref{eq:Frac Dal same space-time order} suggests
a solution of the form, as in the case of integer derivatives:

\begin{equation}
\tilde{\phi}(\xi,\eta)=f(\xi)+g(\eta),\label{eq:proposed solution fi}
\end{equation}
subject to the initial conditions

\begin{equation}
\left\{ \begin{array}{cc}
\phi(x,0)=F(x)\\
{\displaystyle \left.\frac{\partial^{\alpha}\phi(x,t)}{\partial t^{\alpha}}\right|}_{t=0}\equiv G_{*}(x)
\end{array}\right..\label{eq:cond init}
\end{equation}

According to the initial conditions,

\begin{equation}
F(x)=f(x)+g(x),\label{eq:F(x)}
\end{equation}

and applying the fractional derivative of order $\alpha$ to Eq. \eqref{eq:F(x)}
yields:

\begin{equation}
f^{(\alpha)}(x)=F^{(\alpha)}(x)-g^{(\alpha)}(x).\label{eq:Deriv-alfa deF(x)}
\end{equation}

By means of the second of Eqs.\eqref{eq:proposed solution fi}, \eqref{eq:cond init}
and using the chain rule Eq.\eqref{eq:Chainrule nondif func-1}, we
obtain

\begin{equation}
g^{(\alpha)}(x)=\frac{G_{*}(x)}{v^{\alpha}}-(-\mathrm{sgn}(v))^{\alpha}f^{(\alpha)}(x).\label{eq:g-alfa}
\end{equation}

Now assuming $v$ as a positive quantity and substituting Eq. \eqref{eq:Deriv-alfa deF(x)}into
Eq. \eqref{eq:g-alfa}, we can eliminate the $f^{(\alpha)}(x)$ dependence,
resulting in

\begin{equation}
g^{(\alpha)}(x)=\frac{G_{*}(x)}{2}-(-1)^{\alpha}\frac{F^{(\alpha)}(x)}{2}.\label{eq:g-alfa em func de F e G}
\end{equation}

To obtain $g(x)$we fractional integrate the Eq. \eqref{eq:g-alfa em func de F e G},
considering as an initial value problem with initial conditions given
by \eqref{eq:cond init}. This can be done, for example, by applying
Laplace transform and its inverse. The result is

\begin{eqnarray}
g(x) & = & \frac{1}{2\Gamma(\alpha)}{\displaystyle \intop_{0}^{x}\left(x-\tau\right)^{(\alpha-1)}\, G_{*}(\tau)d\tau}+\nonumber \\
 & + & \frac{(-1)^{\alpha}}{2}[F(0)-F(x)]+g(0)\label{eq:g(x) integrado}
\end{eqnarray}

where $\Gamma(\alpha)$ is the gamma function.

We have then found the functional forms of $f$ and $g$, so that
the general solution for a general instant of time, t, can be expressed
as below:

\begin{eqnarray}
\phi\left(\xi,\eta\right) & = & \frac{1}{2\Gamma(\alpha)}{\displaystyle \intop_{\xi}^{0}\left(\xi-\tau\right)^{(\alpha-1)}\, G_{*}(\tau)d\tau}+\nonumber \\
 & + & {\displaystyle \frac{1}{2\Gamma(\alpha)}\intop_{0}^{\eta}\left(\eta-\tau\right)^{(\alpha-1)}\, G_{*}(\tau)d\tau+}\nonumber \\
 & + & F(\xi)[\frac{(-1)^{\alpha}}{2}+1]-F(\eta)\frac{(-1)^{\alpha}}{2}.
\end{eqnarray}

\subsubsection{Analysis of the exponents analysis for preservation of chirality:}

To preserve the chiral symmetry we can choice the fractional exponents
to satisfy $(-1)^{\alpha}=-1.$ This condition can be written as
\begin{equation}
\exp i(\pi+2n\pi)\alpha=\exp i(\pi+2k\pi),n,k\in\left\{ \mathbb{N}\cup\{0\}\right\} ,
\end{equation}
witch leads to condition for the fractional exponents that preserves
chirality as

\begin{eqnarray}
\alpha & = & \frac{2k+1}{2n+1},\qquad0<\alpha<1,\nonumber \\
k & = & 0,1,2...;\qquad n=k+1,k+2...
\end{eqnarray}

\subsubsection{Regularization to preserve chiral symmetry}

In order to impose the preservation of chiral symmetry properties,
we can define an regularized fractional derivative in the MRL sense
as

Definition: ${\partial^{\alpha}}/{\partial\hat{t}^{\alpha}}\equiv\left (\mathrm{sign}(MV_{k})\right)^{\alpha}{\partial^{\alpha}}/{\partial t^{\alpha}},$
where $\mathrm{sign}(MV_{k})$ is the signal o the mover left or right given
by

\begin{equation}
MV_{k}=\left\{ \begin{array}{cc}
-1, & k=\xi\quad \mathrm{ for\: left\: movers}\\
+1, & k=\eta\quad \mathrm{for\;  right\: movers}
\end{array}\right.
\end{equation}

With the above definition, the fractional wave equation could be written
as

\begin{equation}
\frac{\partial^{\alpha}}{\partial x^{\alpha}}\frac{\partial^{\alpha}}{\partial x^{\alpha}}\phi\left(x,t\right)-\frac{1}{v^{2\alpha}}\frac{\partial^{\alpha}}{\partial\hat{t}^{\alpha}}\frac{\partial^{\alpha}}{\partial\hat{t}^{\alpha}}\phi\left(x,t\right)=0
\end{equation}
and the solutions will not carry the complex factor, preserving so
the chirality.

We also proceed similarly with an wave equation with different exponents
in space and time

\begin{equation}
\frac{\partial^{\alpha}}{\partial x^{\alpha}}\frac{\partial^{\alpha}}{\partial x^{\alpha}}\phi\left(x,t\right)-\frac{1}{v^{2\beta}}\frac{\partial^{\beta}}{\partial t^{\beta}}\frac{\partial^{\beta}}{\partial t^{\beta}}\phi\left(x,t\right)=0.\label{eq:Fractional D'alembertian different expon}
\end{equation}

The development is in Appendix A.

The general result can also indicate that a regularized definition
of derivative, similar to the Feller definition by using MRL derivative,
could be more adequate to handle this kind of problem. Also a regularized
definition could give a option to conserve the parity or the chiral
properties of the field.

In the sequence we propose an alternative approach by considering
fractional space-time instead of fractional space functions, that
is, we consider that a coarse-grained space-time, meaning that nor
the space nor the time are infinitely thine but have ''thickness''.

\subsection{Coarse-grained space-time}

We now consider the problem with a coarse-grained space-time which
means that space and time are non-differentiable and considering the
chain rule as \cite{Jumarie1}

\begin{equation}
\frac{d^{\alpha}}{dx^{\alpha}}f[u(x)]=\frac{d}{du}f\,\frac{d^{\alpha}}{dx^{\alpha}}u.\label{eq:chain rule space-time coarse}
\end{equation}

It can be shown that the ansatz $\phi=\phi(x^{\alpha}+\lambda t^{\beta})$
is a solution of the fractional wave equation \eqref{eq:Fractional D'alembertian different expon}
in a coarse-grained space-time \cite{Jumarie1}, subject to the condition

\begin{equation}
\lambda_{\alpha,\beta}=\pm v^{\beta}\frac{\Gamma(\alpha+1)}{\Gamma(\beta+1)}.\label{eq:Lambda versus vbeta}
\end{equation}

The result above gives the insight to redefine the light-cone variables
$\xi,\eta$ as

\begin{equation}
\left\{ \begin{array}{cc}
\xi=x^{\alpha}-\lambda t^{\beta}\\
\eta=x^{\alpha}+\lambda t^{\beta}
\end{array}\right..\label{eq:New Ksi-Eta}
\end{equation}

Applying the rule \eqref{eq:chain rule space-time coarse} to the
fractional D'alembertian \eqref{eq:Fractional D'alembertian different expon}
with the new variables \eqref{eq:New Ksi-Eta}, we obtain, after some
algebra, a simple form

\begin{equation}
{\displaystyle \frac{\partial^{2}}{\partial\xi\partial\eta}\phi\left(\xi,\eta\right)}=0,\label{eq:fractional dalemb-fractional light cone}
\end{equation}
subject to the condition \eqref{eq:Lambda versus vbeta}. Eq. \eqref{eq:fractional dalemb-fractional light cone}
permits to apply the same procedure of subsection \eqref{sub:Solution-into-the light cone}.
The result is

\begin{equation}
\phi\left(\xi,\eta\right)=\frac{1}{2K\Gamma(\beta+1)}{\displaystyle \intop_{\xi}^{\eta}G_{**}(y^{\alpha})dy}+\frac{F(\xi)}{2}+\frac{F(\eta)}{2},
\end{equation}
where
\begin{equation}
{\displaystyle \left.\frac{\partial^{\beta}\phi(x,t)}{\partial t^{\beta}}\right|}_{t=0}\equiv G_{**}(x^{\alpha})
\end{equation}
The advantage of this approach is that there is no violation of chirality
and open the perspective to study higher orders derivatives in fractional
space-time by applying successively the operator given by Eq. \eqref{eq:fractional dalemb-fractional light cone}
to an ansatz $\phi\left(\xi,\eta\right)$.

The introduction of higher derivatives yields the so-called negative
squared-norm states ghosts. Here, we argue that the presence of fractional
higher derivatives might remove the problem of these unphysical modes.

\section{Lorentz transforms and invariance conditions}

The standard Lorentz boosts in(1+1) dimensions read:

\begin{equation}
\left\{ \begin{array}{cc}
x'=\gamma(x-vt)\\
t'=\gamma(-\frac{v}{c^{2}}x+t)
\end{array},\right.\label{eq:Standard Lorentz Transform}
\end{equation}

Its inverse can be obtained by changing v by (-v).

We shall obtain the fractional Lorentz transforms in a way which differs
from Jumarie's approach. Considering now the fractional front wave
as
\begin{eqnarray}
c^{2\alpha}(t{}^{\alpha})^{2}&-&(x^{\alpha})^{2}-(y^{\alpha})^{2}-(z^{\alpha})^{2}= \nonumber \\
c^{2\alpha}(t^{\alpha})^{2}&=&c^{2\alpha}(t'^{\alpha})^{2}-(x'^{\alpha})^{2}-(y'^{\alpha})^{2}-(z'^{\alpha})^{2},\label{eq:fractional front wave}
\end{eqnarray}
we suppose a fractional transformation of form

\begin{equation}
\left\{ \begin{array}{cc}
x'^{\alpha}=\gamma_{\alpha,\beta}(x^{\alpha}-\lambda t^{\beta})\\
t'^{\beta}=\gamma_{\alpha,\beta}(t^{\beta}-\frac{\lambda}{c^{2\beta}}x^{\alpha})
\end{array}.\right.
\end{equation}
The inverse transform can directly be obtained as

\begin{equation}
\left\{ \begin{array}{cc}
x{}^{\alpha}=\gamma_{\alpha,\beta}(x'^{\alpha}+\lambda t'^{\beta})\\
t^{\beta}=\gamma_{\alpha,\beta}(t'^{\beta}+\frac{\lambda}{c^{2\beta}}x^{'\alpha})
\end{array}.\right.\label{eq:InversefracLorentz Transf}
\end{equation}

From Eq.\eqref{eq:fractional front wave}, with the transformations
above, we are lead to a fractional gamma factor which reads as below:

\begin{equation}
\gamma_{\alpha,\beta}=\frac{1}{\sqrt{1-\frac{\lambda_{\alpha,\beta}^{2}}{c^{2\beta}}}}.
\end{equation}

From Eq. \eqref{eq:Lambda versus vbeta}, we see that $\lambda_{\alpha,\beta}$
depends on the fractional exponents $\alpha,\beta.$

\subsection{Fractional Lorentz transform invariance for coarse-grained space-time }

We have shown that a function of $\phi(x^{\alpha},t^{\beta})$ is
a solution of wave equation given by Eq.\eqref{eq:Fractional D'alembertian different expon}.
It can be shown that this wave equation is Invariant to Fractional
Lorentz transform. The details are in Appendix B.

\subsection{Standard Lorentz invariance for non-differentiable space of solutions}

It can also be proved that in the space of non-differentiable solutions,
the fractional wave equation in Lorentz invariant by standard Lorentz
transforms, if exponents of fractional derivatives in space and time
are equal to each o other. The details can also be found in appendix
B.

\section{An explicit example of solution}

As an illustrative example, let us take our initial conditions as
$F(x)=\phi(x,0)=\Theta(x)\Theta(1-x)$, where $\Theta(x)$ is the
Heaviside function and, ${\displaystyle \left.\frac{\partial^{\alpha}\phi(x,t)}{\partial t^{\alpha}}\right|}_{t=0}\equiv G_{*}(x)=0$.
With these conditions, the time evolution of the general solution
for non-differential functions, acquires the form
$F(x)=\phi(x,0)=\Theta(x)\Theta(1-x)$, and ${\displaystyle \left.\frac{\partial^{\alpha}\phi(x,t)}{\partial t^{\alpha}}\right|}_{t=0}\equiv G_{*}(x)=0$.
With these conditions, the time evolution of the general solution
for non-differential functions, acquires the form
\begin{eqnarray}
\phi(x,t)&=& F(x-vt)\left[{{(-1)^{\alpha}}\over{2}}+1\right]-F(x+vt)\left[{{(-1)^{\alpha}}\over{2}}+1\right] \nonumber \\
\phi(x,t)&=& \Theta(x-vt)\Theta(1-x+vt)\left[{{(-1)^{\alpha}}\over{2}}+1\right]+ \nonumber \\
&-&\Theta(x+vt)\Theta(1-x-vt){{(-1)^{\alpha}}\over{2}}.
\end{eqnarray}
The solution above represents two well localized propagating rectangular
pulses, propagating in opposite directions, with different attenuation
parameters that depend on the chirality and the fractional exponent.
Again, if the fractional exponent is one of those that preserves the
chiral symmetry, the solution is identical to the case of an integer
exponent.

For the case of fractional space-time, the solutions is similar for
this example but with different space and time scales with the chiral
symmetry preserved.

\bigskip{}

\section{Concluding Remarks}

In this work, by taking into account a space of non-differentiable
functions in one case and a non-differentiable space-time (coarse-grained)
in the other, we have obtained in a natural way fractional wave equations
in terms of a fractional D'alembertian with a sequential form of modified
fractional Riemann-Liouville. We claim that the novelty of our work
is the particular choice of light-cone coordinates along with the
use of sequential modified fractional derivatives and an adequate
chain rules, leading to technique that creates a perspective to obtain
solutions for other similar problems. Our solutions are worked out
for general initial conditions. The Jumarie's approach of fractional
calculus seems to be more adequate to deal with problems that involve
transformations in coordinates, since the chain and Leibniz rules
are less complicated. Since we are choosing to work with non-differentiable
functions or a coarse-grained space-time, no use of distributional
generalized functions or fractional powers of operators, neither the
maintenance of semi-group properties of exponents in the derivatives
is made. In each case of study the results agrees with standard
integer order in the convenient limits.

In terms of the non-differentiable space of solutions, we have constructed
a wave equation and shown that the space and time orders of fractional
derivatives must be the same in order to make physics sense. A solution
obtained with the light-cone coordinates shows a possible violation
of the chiral separation. The exponents are discriminated in order
to distinguish cases to preserve the chiral properties. A suggestion
to intentionally prevent the chiral violation is presented in terms
of a movers sign regularized fractional derivatives.

In the non-differentiable space-time (coarse grained), the form of
solutions to the wave equation in terms of power of space and time,
gives a path to understand the a fractional Lorentz transform proposed
by Jumarie, in terms of metric invariant radius of a propagating fractional
front wave. The approach used opened up possibilities to further studies
of higher-fractional order wave equations.Complementary, we have explicitly
shown that the fractional wave equation in terms of the non-differentiable
space-time is invariant under a Lorentz transform-like called fractional
Lorentz transform, within the conditions of equality for fractional
orders of derivatives in space and in time. Similar results from fractional
wave equation in a non-differentiable space of solutions functions
in terms of standard Lorentz transform.

We should point out that, if Lorentz symmetry is not at work, the
study of systems with a mismatch between the number of time and space
derivatives is common in a number of condensed-matter systems, as
described in the theory of dynamical critical systems and quantum
criticality \cite{Hohenberg-Halperin,Livro-S.-K.Ma}. The remarkable
aspect of these systems is the presence of scaling properties that
are anisotropic in time and space. Nowadays, the construction of 
gravitational models and field theories is intimately connected
to the number of space-time dimensions, extra dimensions, as well as eventual 
effects of fractional dimensions and possible contributions from non-holonomic
commutative/non-commutative variables of such fractal dimensions.  
We understand that it is important to seek alternative physical concepts by means of 
different approaches of calculus and geometry, considering new ideas and models for 
space-time \cite{Vac}.

More recently, based on the theory
of Lifschitz scalars in arbitrary dimensions \cite{Lifchitz; Hornreich},
Horava introduced a new class of quantum gravity models \cite{Horava}
with the outstanding property of renormalisability in (1+3) dimensions.
Use of FC in this context and related ones is indicated in the Calcagni's
work \cite{Calcagni} and references therein.

\section{Acknowledgments}
The authors J. Weberszpil and J. A.
Helay\"{e}l-Neto wish to express their gratitude to FAPERJ-Rio de Janeiro
and CNPq-Brazil for the partial financial support.

\appendix
\section{Fractional D'Alembertian of different space-time partial derivative
orders}

Eq. \eqref{eq:Fractional D'alembertian different expon} in light-cone
coordinates take the form:

\begin{eqnarray}
\left[\frac{\partial^{\alpha}}{\partial\xi^{\alpha}}\frac{\partial^{\alpha}}{\partial\xi{}^{\alpha}}-{\displaystyle \frac{\partial^{\beta}}{\partial\eta^{\beta}}\frac{\partial^{\beta}}{\partial\eta{}^{\beta}}}\right]\phi\left(\xi,\eta\right) & +\nonumber \\
+2\left[\frac{\partial^{\alpha}}{\partial\xi^{\alpha}}\frac{\partial^{\alpha}}{\partial\eta{}^{\alpha}}-(-1)^{\beta}{\displaystyle \frac{\partial^{\beta}}{\partial\xi^{\beta}}\frac{\partial^{\beta}}{\partial\eta{}^{\beta}}}\right]\phi\left(\xi,\eta\right) & +\nonumber \\
\begin{aligned}+ & \left[\frac{\partial^{\alpha}}{\partial\eta^{\alpha}}\frac{\partial^{\alpha}}{\partial\eta{}^{\alpha}}-{\displaystyle \frac{\partial^{\beta}}{\partial\xi^{\beta}}\frac{\partial^{\beta}}{\partial\xi{}^{\beta}}}\right]\phi\left(\xi,\eta\right)\end{aligned}
 & = & 0.
\end{eqnarray}

The result above is not so compact as the previous one. If $\alpha=\beta\neq1$we
recover the previous result. If $\alpha=\beta=1$, the result is consistent
with the literature, see for example \cite{Equac Diff}.

Now, back to the $\left(t;x\right)$-coordinates, we get to an expression
which is not clearly Eq. \eqref{eq:Fractional D'alembertian different expon}:
\begin{widetext}
\begin{eqnarray}
\left(\frac{1}{2}\right)^{\alpha+\beta}\left[\frac{\partial^{\alpha}}{\partial x^{\alpha}}\frac{\partial^{\beta}}{\partial x^{\beta}}+\frac{1}{v^{\beta}}\frac{\partial^{\alpha}}{\partial x^{\alpha}}\frac{\partial^{\beta}}{\partial t^{\beta}}+(-1)^{\alpha}\frac{1}{v^{\alpha}}\frac{\partial^{\alpha}}{\partial t^{\alpha}}\frac{\partial^{\beta}}{\partial x^{\beta}}+(-1)^{\alpha}\frac{1}{v^{\alpha+\beta}}\frac{\partial^{\alpha}}{\partial t^{\alpha}}\frac{\partial^{\beta}}{\partial t^{\beta}}\right]\phi\left(x,t\right)=0 . \nonumber \\
\label{eq:A2}
\end{eqnarray}
\end{widetext}

Again, if $\alpha=\beta=1$ we recover the traditional D'alembertian;
but, if we transform the above expression back to light-cone of variables,
the result, due to the non-biunivicity of the functions is not the
starting point, ${\displaystyle \frac{\partial^{\alpha}}{\partial\xi^{\alpha}}\frac{\partial^{\beta}}{\partial\eta{}^{\beta}}\phi\left(\xi,\eta\right)}=0$,
but coincide with the literature if $\alpha=\beta=1$. This can be
understood if we imagine fractal space functions and try to go from
one point into to others passing trough a point of ramification, characterizing
the multiplicity of solutions.

Notice that in \eqref{eq:A2}, if we take $\alpha=\beta,$and if $(-1)^{\alpha}=-1,$
chirality is preserved and we recover Eq. \eqref{eq:Fractional D'alembertian different expon}.

\section{Fractional Lorentz transform invariance for coarse-grained space-time}

Using the chain rule \eqref{eq:chain rule space-time coarse} and
the inverse fractional Lorentz transforms given by \eqref{eq:InversefracLorentz Transf},
we obtain

\begin{flalign}
\frac{\partial^{\alpha}\phi}{\partial x'^{\alpha}} & =\frac{\partial\phi}{\partial(x^{\alpha})}\frac{\partial^{\alpha}x^{\alpha}}{\partial x'^{\alpha}}+\frac{\partial\phi}{\partial(t^{\beta})}\frac{\partial^{\alpha}t^{\beta}}{\partial x'^{\alpha}}\nonumber \\
\frac{\partial^{\alpha}\phi}{\partial x'^{\alpha}} & =\frac{\partial\phi}{\partial(x^{\alpha})}\gamma\Gamma(\alpha+1)+\frac{\partial\phi}{\partial(t^{\beta})}\frac{\lambda\gamma\Gamma(\alpha+1)}{c^{2\beta}}.\label{eq:seq1}
\end{flalign}

and using that

\begin{alignat}{1}
\left\{ \begin{array}{cc}
\partial^{\alpha}\phi\cong\Gamma(\alpha+1)\partial\phi\\
\partial^{\beta}\phi\cong\Gamma(\beta+1)\partial\phi
\end{array}\right.,
\end{alignat}
we can re-write Eq. \eqref{eq:seq1} as

\begin{equation}
\frac{\partial^{\alpha}\phi}{\partial x'^{\alpha}}=\gamma\frac{\partial^{\alpha}\phi}{\partial x^{\alpha}}+\frac{\lambda\gamma\Gamma(\alpha+1)}{c^{2\beta}\Gamma(\beta+1)}\frac{\partial^{\beta}\phi}{\partial t^{\beta}}.
\end{equation}

Analogously, we can write for the fractional temporal derivative

\begin{equation}
\frac{\partial^{\beta}\phi}{\partial t'^{\beta}}=\gamma\frac{\partial^{\beta}\phi}{\partial t^{\beta}}+\frac{\lambda\gamma\Gamma(\beta+1)}{\Gamma(\alpha+1)}\frac{\partial^{\alpha}\phi}{\partial x^{\alpha}}.
\end{equation}

Repeating this procedure for successive factional derivatives, and
assuming as a premise that the fractional wave equation is valid in
the S referential, we show that it is valid in the S' referential
as follows after some rearrangements

\begin{widetext}
\begin{eqnarray}
\label{eq:B5}
\frac{\partial^{\alpha}}{\partial x'^{\alpha}}\frac{\partial^{\alpha}\phi}{\partial x'^{\alpha}}-\frac{1}{v^{2\beta}}\frac{\partial^{\beta}}{\partial t'^{\beta}}\frac{\partial^{\beta}\phi}{\partial t'^{\beta}}&=&\gamma^{2}\left[\frac{\partial^{\alpha}}{\partial x^{\alpha}}\frac{\partial^{\alpha}\phi}{\partial x^{\alpha}}-\frac{1}{v^{2\beta}}\frac{\partial^{\beta}}{\partial t^{\beta}}\frac{\partial^{\beta}\phi}{\partial t^{\beta}}\right]
+\frac{2\lambda\gamma^{2}\Gamma(\alpha+1)}{c^{2\beta}\Gamma(\beta+1)}\frac{\partial^{\alpha}}{\partial x^{\alpha}}\frac{\partial^{\beta}\phi}{\partial t^{\beta}}+\nonumber \\
&-&\frac{2\lambda\gamma^{2}\Gamma(\alpha+1)}{v^{2\beta}\Gamma(\beta+1)}\frac{\partial^{\alpha}}{\partial x^{\alpha}}\frac{\partial^{\beta}\phi}{\partial t^{\beta}}-\left[-\frac{\lambda^{2}\gamma^{2}\Gamma^{2}(\alpha+1)}{c^{4\beta}\Gamma^{2}(\beta+1)}\frac{\partial^{\beta}}{\partial t^{\beta}}\frac{\partial^{\beta}\phi}{\partial t^{\beta}}+\frac{\lambda^{2}\gamma^{2}\Gamma^{2}(\beta+1)}{v^{2\beta}\Gamma^{2}(\alpha+1)}\frac{\partial^{\alpha}}{\partial x^{\alpha}}\frac{\partial^{\alpha}\phi}{\partial x^{\alpha}}\right]. \nonumber\\
\label{eq:B5}
\end{eqnarray}
\end{widetext}

Using that $\alpha=\beta$ and $v^{\beta}=c^{\beta},$ we obtain that
the LHS of \eqref{eq:B5} is zero, that is, invariant for fractional
Lorentz transforms.

\subsection*{Standard Lorentz invariance for non-differentiable space of solutions}

Using the inverse of standard Lorentz transform \eqref{eq:Standard Lorentz Transform}
and the chain rule for non-differentiable functions \eqref{eq:Chainrule nondif func-1}
we can write

\begin{eqnarray}
\frac{\partial^{\alpha}\phi}{\partial x'^{\alpha}} & = & \frac{\partial^{\alpha}\phi}{\partial x^{\alpha}}\left(\frac{\partial x}{\partial x'}\right)^{\alpha}+\frac{\partial^{\alpha}\phi}{\partial t^{\alpha}}\left(\frac{\partial t}{\partial x'}\right)^{\alpha}=\nonumber \\
 & = & \frac{\partial^{\alpha}\phi}{\partial x^{\alpha}}\gamma^{\alpha}+\frac{\partial^{\alpha}\phi}{\partial t^{\alpha}}\left(\gamma\frac{v}{c}\right)^{\alpha},
\end{eqnarray}

\begin{eqnarray}
\frac{\partial^{\beta}\phi}{\partial t'^{\beta}} & = & \frac{\partial^{\beta}\phi}{\partial x^{\beta}}\left(\frac{\partial x}{\partial t'}\right)^{\beta}+\frac{\partial^{\beta}\phi}{\partial t^{\beta}}\left(\frac{\partial t}{\partial t'}\right)^{\beta}=\nonumber \\
 & = & \frac{\partial^{\alpha}\phi}{\partial x^{\alpha}}\left(\gamma v\right)^{\alpha}+\frac{\partial^{\alpha}\phi}{\partial t^{\alpha}}\gamma^{\alpha}.
\end{eqnarray}

Proceeding again with the derivatives and assuming as a premise the
validity of Eq.\eqref{eq:Fractional D'alembertian different expon},
we obtain that the fractional wave equations, in a space of non-differentiable
solutions is Standard Lorentz invariant, as can be seen below
\begin{widetext}
\begin{eqnarray}
\frac{\partial^{\alpha}}{\partial x'^{\alpha}}\frac{\partial^{\alpha}\phi}{\partial x'^{\alpha}}-\frac{1}{c^{2\beta}}\frac{\partial^{\beta}}{\partial t'^{\beta}}\frac{\partial^{\beta}\phi}{\partial t'^{\beta}}
=\gamma^{2\alpha}\left[\frac{\partial^{\alpha}}{\partial x^{\alpha}}\frac{\partial^{\alpha}\phi}{\partial x^{\alpha}}-\frac{1}{c^{2\beta}}\frac{\partial^{\beta}}{\partial t^{\beta}}\frac{\partial^{\beta}\phi}{\partial t^{\beta}}\right] +\frac{2\gamma^{2\alpha}v^{\alpha}}{c^{2\alpha}}\frac{\partial^{\alpha}}{\partial x^{\alpha}}\frac{\partial^{\alpha}\phi}{\partial t^{\alpha}}-\frac{2\gamma^{2\beta}v^{\beta}}{c^{2\beta}}\frac{\partial^{\beta}}{\partial x^{\beta}}\frac{\partial^{\beta}\phi}{\partial t^{\beta}}&+ \nonumber \\
-\left[-v^{2\alpha}\gamma^{2\alpha}\frac{\partial^{\alpha}}{\partial t^{\alpha}}\frac{\partial^{\alpha}\phi}{\partial t^{\alpha}}+\gamma^{2\beta}\frac{v^{2\beta}}{c^{2\beta}}\frac{\partial^{\beta}}{\partial x^{\beta}}\frac{\partial^{\beta}\phi}{\partial x^{\beta}}\right] &  & ,
\end{eqnarray}
\end{widetext}
that is identically zero if $\alpha=\beta$ and $v^{\beta}=c^{\beta}$.

\end{document}